\documentclass[twocolumn,showpacs,preprintnumbers,amsmath,amssymb]{revtex4}

\usepackage{color}

\usepackage{graphics}
\usepackage{graphicx}
\usepackage{dcolumn}
\usepackage{bm}
\usepackage{amssymb}
\begin{document}
\preprint{APS/123-QED}

\title{Magneto-elastic coupling in the spin-Peierls ground state of hydrogenated and deuterated (TMTTF)$_2$PF$_6$ salts}
\author{Mario Poirier$^1$}
\author{Alexandre Langlois$^1$}
\author{Claude Bourbonnais$^{1,2}$}
\affiliation{$^1$Regroupement Qu\'eb\'ecois sur les Mat\'eriaux de
Pointe, D\'epartement de Physique, Universit\'e de Sherbrooke,
Sherbrooke, Qu\'ebec,Canada J1K 2R1}
\affiliation{$^2$Canadian Institute for Advanced Research, Toronto, Canada}
\author{Pascale Foury-Leylekian}
\author{Alec Moradpour}
\author{Jean-Paul Pouget}
\affiliation{ Laboratoire de Physique des Solides, CNRS UMR 8502,
Universit\'e Paris-Sud, 91405 Orsay C\'edex, France}

\date{\today}

\begin{abstract}
We report an ultrasonic study of the magneto-elastic coupling of the hydrogenated and deuterated (TMTTF)$_2$PF$_6$ organic salts. For both salts the temperature dependence of the longitudinal velocity along the ${\textit{\textbf{c}}}^*$ axis displays a monotonic stiffening of the $C_{33}$ compressibility modulus upon cooling. Below the characteristic temperature scale 40 K the modulus stiffening becomes markedly enhanced, in concomitance with the reduction of spin degrees of freedom  previously seen in magnetic measurements as low dimensional precursors of the spin-Peierls transition. The magneto-elastic coupling appears to be much weaker in the hydrogenated salt due to the highly inhomogeneous elastic behavior induced by the proximity of the charge ordering transition  to the spin-Peierls phase. For the deuterated salt, an important  anomaly in the ultrasound velocity is observed below the spin-Peierls transition temperature $T_{\rm SP}$ in agreement with scaling of the elastic deformation with the spin-Peierls order parameter. In spite of the weakly inhomogeneous character of the spin-Peierls phase transition, the magnetic field dependence of $T_{\rm SP}$ is well captured  with the mean-field prediction for the lattice distorted Heisenberg spin chain.

\end{abstract}

\maketitle
\section{INTRODUCTION}
The sulfur series of quasi-one-dimensional (1D) organic compounds (TMTTF)$_2 X$ with the centro-symmetrical anions $X=$ PF$_6$, AsF$_6$, SbF$_6$... are Mott insulators which develop a charge ordered (CO) state \cite{Bourbon2008,Brown2008,Monceau2008} that is followed at lower temperature by either an antiferromagnetic (AF) N\'eel or a lattice distorted spin-Peierls (SP) phase depending on the pressure applied, anion $X$  substitution, and to some degree  the  deuteration of the methyl groups. In low pressure conditions, these (TMTTF)$_2 X$ salts stand out as model low-dimensional systems to study the competition between spin, charge, and lattice degrees of freedom.

The (TMTTF)$_2$PF$_6$ compound is particularly well positioned in the phase diagram allowing a systematic study of the interplay between differing degrees of freedom and the mechanism involved in the stabilization of a variety of phases. Indeed, this compound undergoes a SP transition at ambient pressure followed by an AF ground state above 9 kbar \cite{Creuzet1985,Caron88,Chow98}. As regards to spin degrees of freedom under normal pressure conditions, the spin singlet state associated with the SP lattice distortion has been borne out by spin susceptibility \cite{Creuzet1987,Dumm00,Foury08}, and NMR spin relaxation rate \cite{Creuzet1987,Wzietek1993,Zamborszky2002B}, whereas a sizable part of the magnetic field-temperature phase diagram for the PF$_6$ salt has been obtained by high-field NMR and magnetization studies \cite{Brown1998}. In the  hydrogenated (H$_{\rm 12}$) PF$_6$ salt, the SP transition takes place at $T_{\rm SP}
\approx 18$ K \cite{Creuzet1987,Foury04,Chow98} and at 13 K in the deuterated (D$_{\rm 12}$) compound \cite{Pouget2006}.  In the H$_{\rm 12}$-PF$_6$ salt, the observation of X-ray 2\textit{$k_F$} diffuse scattering indicates the presence  below $\sim$ 60 K  of one-dimensional lattice precursors of the SP transition \cite{Pouget1982}; these eventually lead to a reduction of spin degrees of freedom akin to a spin pseudogap \cite{Bourbonnais96}, as exhibited by EPR susceptibility \cite{Creuzet1987,Dumm00}, ${^1}\!$H \cite{Creuzet1985,Chow98} and $^{13}\!$C \cite{Creuzet1987} NMR data.

If the role of lattice degrees of freedom in the stabilization of the CO ground state is relatively well documented in the PF$_6$ and AsF$_6$ salts \cite{deSouza2008,deSouza2009}, the situation is not so clear for the SP transition. Except for  X-ray diffuse scattering for the lattice precursors below 60 K, which condense into Bragg spots at $T_{\rm SP}$ \cite{Pouget1982}, and anomalies in the thermal expansivity data \cite{deSouza2008,deSouza2009}, the nature of the magneto-elastic coupling involved in the SP instability remains poorly understood in these compounds.  As far as the charge degrees of freedom are concerned, they appear largely decoupled from the spins in the non ordered temperature region. At $T_{\rm SP}$ however, a small dielectric anomaly has been observed in both H$_{\rm 12}$ and D$_{\rm 12}$ PF$_6$ salts, indicating a finite interaction between them as a result of three-dimensional ordering \cite{Langlois2010}.

In this paper, we use the ultrasound technique to investigate magneto-elastic coupling effects on the ${\textit{\textbf{c}}}^*$ axis compressibility modulus of the H$_{{12}}$ and D$_{{12}}$ (TMTTF)$_2$PF$_6$ salts. Our data reveal the highly inhomogenous character of elastic properties of the H$_{\rm 12}$ salt, compared to the D$_{\rm 12}$ one, which is ascribed to the proximity of the CO and SP phases in the H$_{\rm 12}$ salt. This inhomogeneous phase yields in turn weaker magneto-elastic effects at low temperature when approaching the SP transition. Two anomalies are observed on the compressibility modulus as the temperature is decreased: from 40 K, the modulus stiffening displays a marked  enhancement, which is superimposed at $T_{\rm SP}$ by a sharper increase, as a result of the SP order parameter. For the D$_{\rm 12}$  salt, precursor effects to the true 3D ordering appear at  $T_{\rm SP}^* \approx$ 15 K and are found to be inhomogeneous in character. The phase diagram obtained  up to a 18 Tesla magnetic field is found to agree with previous measurements \cite{Langlois2010,Brown1998} and the mean-field prediction for the lattice distorted Heisenberg spin chain under field \cite{Cross1979}.

\section{EXPERIMENT}
Hydrogenated (H$_{\rm 12}$) (TMTTF)$_2$PF$_6$ single crystals were grown from THF using the standard constant-current (low current density) electrochemical procedure.  Starting from a deuterated chloroketone \cite{Langlois2010}, the synthesis of the fully deuterated D$_{\rm 12}$ was obtained using usual procedures and the corresponding deuterium incorporation-levels is estimated by NMR to be 97.5 \%. The crystals grow as needles oriented along the chain axis $\textit{\textbf{a}}$ and with two natural parallel faces perpendicular to the ${\textit{\textbf{c}}}^*$ axis. We use a pulsed ultrasonic interferometer to measure the variation of the longitudinal acoustic velocity along ${\textit{\textbf{c}}}^*$ relative to its value at $T_0$ = 30 K, $\Delta V_L / V_L$ = $[V_L(T) - V_L(T_0)] / V_L(T_0)$. The acoustic pulses are generated with LiNbO$_3$ piezoelectric transducers resonating at 30 MHz and odd overtones bonded to the crystals with silicone seal. Since the crystal structure is triclinic, the acoustic mode is characterized as quasi-longitudinal and is dominated by the $C_{33}$ elastic constant or the ${\textit{\textbf{c}}}^*$-axis compressibility modulus. The ultrasonic technique is used in the transmission mode and, because the typical thickness of the crystal along the ${\textit{\textbf{c}}}^*$-axis is quite small around 0.4 mm, a CaF$_2$ delay line must be used to separate the first transmitted acoustic echo from the electric pulse. Moreover, no transverse acoustic mode can be properly analyzed because of mode conversion and mode mixing at the different interfaces. The longitudinal mode can be measured because it has the largest velocity that permits its time separation from parasitic signals. Since the velocity is related to the mass density $\rho$ and the $C_{33}$ modulus through the relation $C_{33} = \rho V_L^2$, the $\Delta V_L / V_L$ data are directly the image of the relative variation of the compressibility modulus $C_{33}$ if the density variations can be ignored. Indeed, thermal expansion measurements \cite{deSouza2008} predicts relative density variation values one to two orders of magnitude smaller than the $\Delta V_L / V_L$ data reported in this paper. Several crystals from different growth batches were used for these ultrasonic experiments. All crystals of the same compound showed similar overall temperature and magnetic field behaviors. The temperature was varied between 2 and 140 K and a magnetic field up to 18 Tesla could be applied along the ${\textit{\textbf{c}}}^*$-axis. Unless otherwise stated in the discussion, all velocity data were obtained for a frequency around 100 MHz.

\begin{figure}[H,h]
\includegraphics[width=8.5cm]{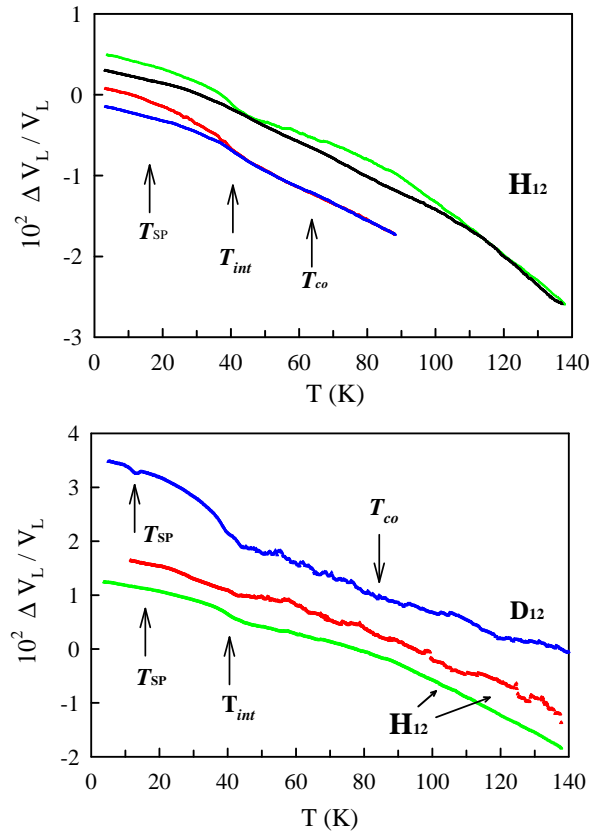} \caption{(Color online)
Temperature dependence of the relative variation of the longitudinal velocity $\Delta V_L / V_L$. Upper panel PF$_6$ (H$_{\rm 12}$): first (red) and third (blue) warming up to 90 K; 24 hours relaxation at 3 K followed by heating up to 140 K (green) and cooling to 2 K (black). Lower panel PF$_6$ ({D\tiny{12}}) and PF$_6$ (H$_{\rm 12}$): cooling down from 140 K (blue and red) and warming to 140 K (green). The curves have been shifted vertically for comparison and the green curves are the same in both panels. Arrows indicate the transition temperatures defined in ref.\cite{deSouza2008}.}\label{fig.1}
\end{figure}

\section{RESULTS AND DISCUSSION}

We present in Figure~1 the temperature dependence of $\Delta V_L / V_L$ below 140 K for the two compounds. In the lower panel, the two top curves (blue and red) refer to the first cooling down of the D$_{\rm 12}$ and the H$_{\rm 12}$ crystals, respectively. At relatively high temperature the curves are noisy owing to the typical degradation of the interfaces between the organic crystal, the piezoelectric transducer and the CaF$_2$ delay line upon cooling due to very different thermal expansion coefficients. Despite this degradation, we can identify two clear features for the D$_{\rm 12}$ compound: beyond the typical velocity stiffening generally observed in solids as the temperature is decreased, there is an extra enhancement in the stiffening below the temperature scale $T_{int}$ $\sim$ 40 K noted in concordance with the characteristic feature seen in thermal expansion data \cite{deSouza2008}. It is followed by another anomaly at much lower temperatures near $T_{\rm SP}$.

For the H$_{\rm 12}$ compound, the same feature is present at the same  $T_{int}$, but with a much smaller amplitude. However, it can be enhanced by warming the sample well above 40 K, that is in a temperature range where the thermal expansion coefficient along the ${\textit{\textbf{c}}}^*$-axis shows the strongest variation \cite{deSouza2008}. This process has yielded the green curve of the lower panel in Fig.1  where the enhancement below $T_{int}$ $\sim$ 40 K sharpens significantly, contrasting with the virtual absence of anomaly at the true transition $T_{\rm SP}$.

Thus for the H$_{\rm 12}$ compound, not only are we observing much smaller elastic anomalies, but it appears that the system enters a highly inhomogeneous elastic phase below 100 K as it can be observed in the upper panel of Figure~1 for different thermal cycles. When the temperature is maintained below 90 K, warming up the crystal several times produces a depression of $\Delta V_L / V_L$ below $T_{int}$ (red and blue curves, upper panel of Fig.~1). To retrieve the velocity enhancement below 40 K, we need first to relax the sample at 3 K during 24 hours, then to heat it up to 140 K (green curve, upper panel of Fig.~1); subsequent cooling down to 2 K (black curve, upper panel  of Fig.~1) reduces the enhancement below 40 K as observed during the first cooling down (red curve, lower panel). However, we observe then a separation of the two curves below 100 K, which is indicative of an inhomogeneous state. Such a behavior is not found in the D$_{\rm 12}$ compound over the same temperature range: repeated thermal cycles to 40 K yield always the same temperature dependence and, although only one other thermal cycling to 200 K (in a magnetic field of 14 Tesla) has been performed, no measurable inhomogeneity effects were detected. Below the crystallization transition near 200 K, the silicone bonds used with organic compounds are highly stable and reproducible and, thus, the inhomogeneous conditions observed here are intrinsic to the samples. They can be possibly linked with the closer proximity of the CO and SP orderings occurring at $T_{\rm CO} = 63$ K \cite{Langlois2010,Nad2000} and $T_{\rm SP}=16.5$ K, respectively in the H$_{\rm 12}$ compound, compared to the $T_{\rm CO} = 84$ K and $T_{\rm SP}= 13$ K in the D$_{\rm 12}$ \cite{Langlois2010}.

As to the nature of degrees of freedom that can be held responsible for this elastic behavior, those of the charge sector are unlikely to be involved, at the very least directly since, according to electrical transport, charge excitations are known to be thermally activated up to relatively high temperature \cite{Coulon82}. However, a softening of the Young's modulus was reported around the CO transition in other (TMTTF)$_2$X salts, the amount of softening being 2\% for AsF$_6$, 0.5\% for SbF$_6$, and 3\% for ReO$_4$ \cite{Brown1989}. This could be related to the charge disproportionation between adjacent TMTTF molecules to which the Young's modulus, dominated by the $C_{11}$ elastic constant along the chain axis, could be more sensitive. For the $c^*$ axis, the $C_{33}$ modulus rather probes the deformation between the organic and anion planes that remains essentially unaffected by the charge redistribution. This may explain why no similar elastic anomaly is detected on either D$_{\rm 12}$ or H$_{\rm 12}$ sample at the charge ordering scale, $T_{\rm CO}$, in the insulating state (see Figure~1).

As for spin degrees of freedom, NMR and spin susceptibility data display essentially gapless excitations in this sector \cite{Creuzet1987,Dumm00,Wzietek1993}, down to the temperature scale of 60 K where, according to early X-ray measurements, lattice precursors of the SP transition appear \cite{Pouget1982}, and  below which spins and lattice become strongly coupled. Spins are known to couple easily to strain in various magnetic systems so as to induce a monotonic softening on the compressibility modulus whose temperature dependence correlates with the regular increase of  magnetic susceptibility with temperature above the transition temperature. For weakly coupled insulating spin chains for example \cite{Trudeau1992,Poirier1995}, the softening upon warming presents a maximum  at a temperature comparable to the spin exchange constant namely, $T \approx J / k_B$, in agreement with the characteristic temperature for the maximum in the temperature dependent spin susceptibility \cite{Dumoulin1997}. The situation differs slightly for the (TMTTF)$_2$PF$_6$ salts considered here. The insulating gap is actually relatively small \cite{Coulon82,Laversanne84} and the spins weakly localized \cite{Bourbon2008}, so that the maximum in susceptibility is pushed above room temperature \cite{Creuzet1987,Dumm00} and so will be the maximum of the elastic softening. Thus the monotonic decay of spin susceptibility appears consistent with the regular increase of sound velocity with decreasing temperature below 140 K (see Fig.~1). Moreover, the upturn in elastic stiffening below $T_{int}$ $\approx$ 40 K corresponds to a decline in magnetic susceptibility and NMR spin-lattice relaxation rate \cite{Creuzet1985,Creuzet1987,Chow98,Foury08}, which is indicative of the lost of spin degrees of freedom due to lattice fluctuations \cite{Bourbonnais96,Pouget1982}. The interpretation given here for $T_{int}$ demarcates from the one given in the framework of thermal expansion experiments \cite{deSouza2008,deSouza2009}, which associated $T_{int}$ with different CO patterns or ferroelectricity.

Let us now examine the magneto-elastic coupling at lower temperature, namely in the vicinity of the SP transition. Because of the highly inhomogeneous elastic character of the H$_{\rm 12}$ compound, we can only discuss the data for the D$_{\rm 12}$ compound shown in Figure~2. Below 40 K, the $\Delta V_L / V_L$ velocity data increases smoothly, presents a local maximum near $T_{\rm SP}^* \approx$ 15 K and then decreases irregularly down to a minimum value at $T_{\rm SP}$ = 13 K. In the SP ordered state, $\Delta V_L / V_L$ starts increasing  again with a temperature dependence reminiscent of the onset of an order parameter tied to a continuous phase transition.

\begin{figure}[H,h]
\includegraphics[width=8.5cm]{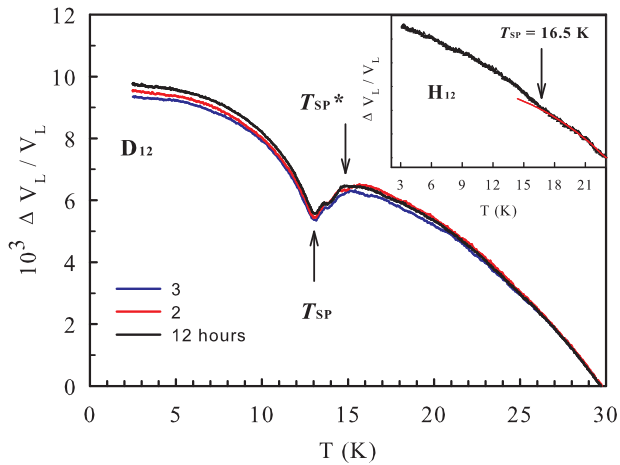} \caption{(Color online)
Temperature dependence of the relative variation of the longitudinal velocity $\Delta V_L / V_L$ for the D$_{\rm 12}$ compound in the SP ground state below 30 K (warming cycle). The plots correspond to different relaxation periods at 3 K. Inset: data in zero field for the H$_{\rm 12}$ compound where the small anomaly at $T_{\rm SP}$ is indicated by an arrow, the red curve being a guide to the eye.} \label{fig.2}
\end{figure}

Although the velocity data do not indicate an inhomogeneous character in the temperature range above  $T_{\rm SP}$, we have verified the stability of the elastic behavior against relaxation effects in the vicinity of the SP transition. The curves shown in Figure~2 were all obtained in warming cycles from 2 K to 40 K, after different relaxation periods at 3 K. Only small differences in amplitude are found below $\sim$ 20 K, but as regards the values of $T_{\rm SP}$ and $T_{\rm SP}^*$ both remain unaffected. The existence of hysteresis effects up to 20 K indicates that the system enters into some kind of inhomogeneous short-range ordered phase, which is distinct from to true SP transition located at $T_{\rm SP}$ = 13 K and where the ultrasound velocity exhibits its characteristic upturn. The low temperature limit attained by the velocity is found to be relaxation period dependent (see Fig.~2). This inhomogeneous character linked to the SP phase was also observed in the microwave dielectric measurements on the D$_{\rm 12}$ salt but was absent for the H$_{\rm 12}$ one \cite{Langlois2010}. In fact, the above elastic features tied to the SP ordering are much less pronounced for the H$_{\rm 12}$ salt: a stiffening anomaly at $T_{\rm SP}\approx 16.5$ K is scarcely visible through a change of slope in the ultrasound velocity (see inset of Fig.~2). Because of the highly inhomogeneous elastic character below 40 K, this tiny anomaly is only observed during the first cooling down from 140 K and no hysteresis precursors appear to be observed above the transition at $T_{\rm SP}$. It appears then that the elastic inhomogeneous character proceeds from two sources in these compounds: the outcome of a local electric polarization tied at high temperature to the spin degrees of freedom affecting the H$_{\rm 12}$ salt over the full temperature range below 100 K due to the proximity of the CO and SP phases, and a short-range SP order below 20 K preceding the true SP transition for the D$_{\rm 12}$ compound. Neutron scattering experiments have revealed a weak three-dimensional (3D) SP amplitude in these salts \cite{Foury04,Pouget2006} likely originating from disorder in the 3D coupling between the chains; as this coupling proceeds from the PF$_6$ anion displacement, any inhomogeneities in the CO ordering  will produce 3D disorder that may be the source of the inhomogeneous SP phase.

\begin{figure}[H,h]
\includegraphics[width=8.5cm]{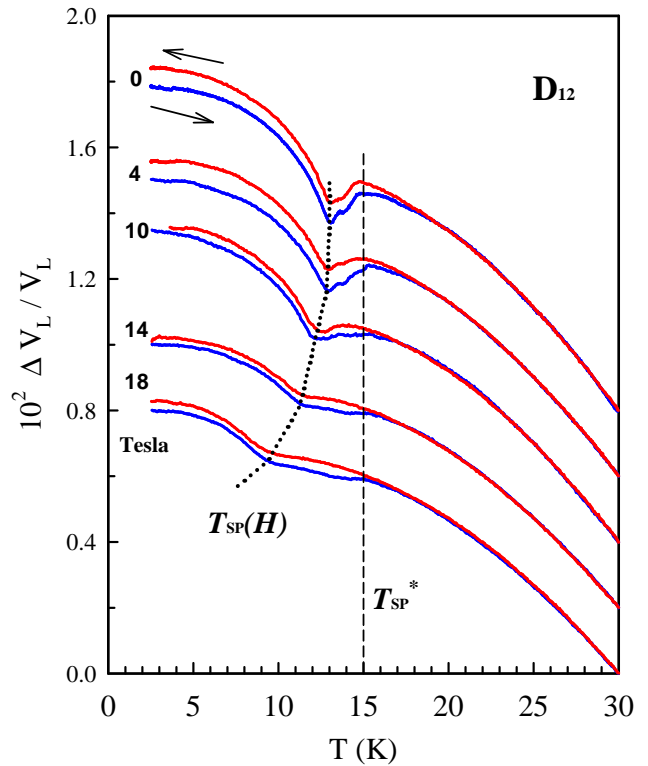} \caption{(Color online)
Temperature dependence of the relative variation of the longitudinal velocity $\Delta V_L / V_L$ for the D$_{\rm 12}$ compound in the SP ground state below 30 K at different magnetic field values. Arrows indicate the variation of temperature. The dotted line is a guide to the eye for the field variation of $T_{\rm SP}(H)$ and the dashed line indicates $T_{sp}^*$ below which an inhomogeneous SP phase appears.}
\label{fig.3}
\end{figure}

To go further into the analysis of the SP state, we present in Figure~3 the magnetic field dependence of the ultrasound velocity for the D$_{\rm 12}$ compound below 40 K. Because of the above mentioned relaxation effects in this temperature range, all the curves have been obtained after a stabilization period of 23 hours at 3 K before warming the crystal up to 40 K (blue curve), followed by a cooling down to 2 K (red curve). The inhomogeneous character of the SP transition is clearly revealed upon warming (blue curves) and cooling (red curves) at all magnetic field values; the curves are clearly shifted from one another below 20 K and more particularly below $T_{\rm SP}^*$, as indicated by the dashed line. When the magnetic field is progressively increased, $T_{\rm SP}$ decreases as expected for a SP transition \cite{Cross1979}, but identically for both thermal cycles, which is consistent with the second order character of the SP transition. The situation is different for $T_{\rm SP}^*$, which is field independent for the warming cycle, but decreases rapidly for the cooling one so as to merge rapidly with $T_{\rm SP}$. Such an  hysteresis loop around  $T_{\rm SP}^*$ is consistent with short range correlations or precursory effects that are progressively accentuated by the magnetic field. Similar effects could not be found in the {H$_{\rm 12}$} compound for which the magneto-elastic coupling appears to be much smaller.

As mentioned before, the $\Delta V_L / V_L$ data in the SP ground state mimics the temperature dependence of an order parameter. Indeed, elastic anomalies at a Peierls or spin-Peierls transition can be simply accounted for by a very simple Landau theory \cite{Pouget1989}, using two order parameters, the modulation (of wave vector \textbf{q}) proportional to the SP magnetic gap $\Delta_q$ and the uniform elastic deformation $e$. In addition to the SP condensate $F_{sp} = a|\Delta_q|^2 + b|\Delta_q|^4$ and the elastic $F_{el} = C_{33}^o e^2/2$ energies, the coupling energy $F_c = h |\Delta_q|^2 e^r$ determines the type of elastic anomaly appearing at the phase transition, $h$ being the coupling constant. Translational symmetry imposes the quadratic $|\Delta_q|^2$ dependence; the term with $r$=2 is always allowed by symmetry, but $r$=1 implies inequivalence of $|e|$ and -$|e|$ elastic deformations. When minimizing the free energy with respect to the elastic deformation $e$, one can show that the biquadratic ($r$=2) coupling term leads to a stiffening of the compressibility modulus $C_{33}$ according to  the expression,
\begin{equation}
C_{33} = C_{33}^o + 2 h |\Delta_q|^2,
\end{equation}
where $C_{33}^o$ is the bare modulus.

On the other hand, to obtain a sudden softening at  a transition (like at $T^*_{\rm SP}$ for instance), a linear to quadratic coupling ($r$=1) is needed, which yields
\begin{equation}
C_{33} = C_{33}^o - \frac{h^2}{2 b},
\end{equation}
where $b$ is the coefficient of the quartic term $|\Delta_q|^4$.

\begin{figure}[H,h]
\includegraphics[width=8.5cm]{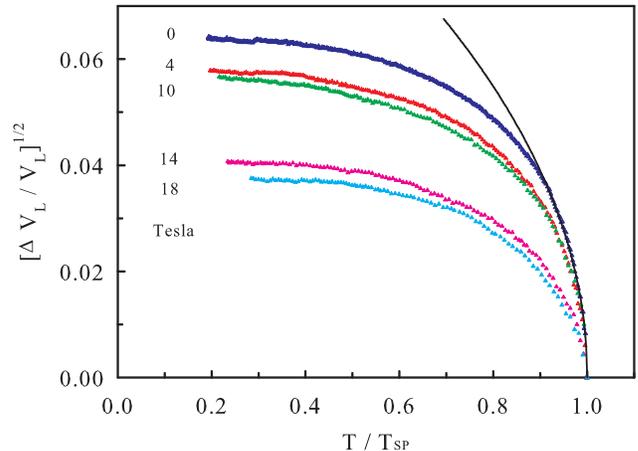} \caption{(Color online)
Temperature dependence of the square root of the relative variation of the longitudinal velocity $\Delta V_L / V_L$  as a function of the reduced temperature at fixed magnetic field values. The full line is a fit to $(1 - T/T_{\rm SP})^\beta$ ($\beta$ = 0.49(2)).}
\label{fig.5}
\end{figure}

We show in Figure~4 the square root of the velocity stiffening below $T_{\rm SP}$, which is proportional to $|\Delta_q|$, as a function of the reduced temperature $T/T_{\rm SP}$. The amplitude of the stiffening anomaly does seem to decrease monotonically with increasing field: according to Eq.~1, this means that either the SP gap $|\Delta_q|$ or the coupling constant $h$ decreases with increasing field. For other SP materials, $|\Delta_q|$ is not considered to be the singlet-triplet gap but a field independent energy required to break down a dimer in the SP ground state \cite{Poirier1995,Azzouz1996}. If this is true for these organic salts, this means that the coupling constant is decreasing with field, $h(H)$. Near the transition temperature the order parameter varies as $\Delta_q \sim (1 - T/T_{\rm SP})^\beta$, with a mean field exponent $\beta \simeq 0.49(2)$ in zero field, as indicated by the black curve in Figure~4. The exponent increases weakly with magnetic field. This mean-field value contrasts with the $\beta \simeq 0.36$ obtained for the dielectric constant enhancement $\Delta\epsilon'$, measured on the hydrogenated compound, which was rather consistent with non classical exponent for a -- one component -- SP order parameter in three dimensions \cite{Langlois2010}. However, pre-transitional 3D fluctuations are observed in the dielectric measurements, while they are practically absent in the elastic ones. Finally, the softening of the velocity data observed below $T_{\rm SP}^*$ down to $T_{\rm SP}$ is rather consistent with Eq.~2 and we must then consider the possibility that it may be dynamically enhanced; as no frequency dependence effects have been investigated, we leave this as an open question. Since $T_{\rm SP}^*$ is not dependent on field, they can be considered as an energy precursor $|\Delta_q^*|$ in an inhomogeneous SP phase.

\begin{figure}[H,h]
\includegraphics[width=8.5cm]{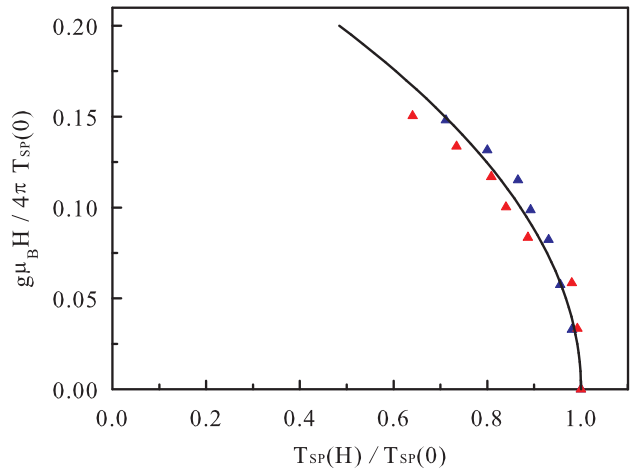} \caption{(Color
online) Magnetic field dependence of the reduced SP transition
temperature $T_{\rm SP}(H)/T_{\rm SP}(0)$ for the deuterated (TMTTF)$_2$PF$_6$ salt: red triangles and blue squares are associated to different criteria defining $T_{\rm SP}(H)$; the continuous line is the best fit obtained from Ref.19.} \label{fig.4}
\end{figure}

The decay of the SP transition temperature under magnetic field  is known to fit a quadratic variation  in relatively low fields  \cite{Bloch1980,Poirier1995,Langlois2010}, which conforms  with  the mean-field prediction for a quasi-one-dimensional SP system \cite{Cross1979,Azzouz1996}. From field dependent dielectric measurements at microwave frequencies \cite{Langlois2010}, it was established that the SP transition temperature for both the {D$_{\rm 12}$} and {H$_{\rm 12}$} salts follows the mean-field expression obtained by Cross \cite{Cross1979},
\begin{equation}
\frac{T_{\rm SP}(H)}{T_{\rm SP}(0)} = \left[1 - c{\left(
\frac{g\mu_BH}{ 4\pi T_{\rm SP}(0)}\right)}^2\right],
\end{equation}
where the $g$-factor is fixed at 2 for (TMTTF)$_2$PF$_6$ \cite{Coulon82}, and the predicted value for the constant $c$ is $14.4$ \cite{Cross1979}. Because of the inhomogeneous character of the SP ground state shown in Figure~3, it is difficult to define only one criterion for $T_{\rm SP}(H)$. In Figure~5 we present the field dependence by using the 1$^{st}$ (blue) and the 2$^{nd}$ (red) temperature derivatives, the 2$^{nd}$ derivative corresponding to the minimum of $\Delta V_L / V_L$ (dotted line in Fig.~3). We have added the fit to Eq.~3 with $c$ = 12.9 deduced from microwave dielectric data obtained on crystals of the same batch \cite{Langlois2010}. The agreement is good, but the dispersion of data  points yields a value $c$ between 10 and 15 due to the presence of the SP inhomogeneous phase. The fair agreement with the quasi-one-dimensional SP mean field prediction contrasts with the situation encountered in other SP systems like CuGeO$_3$ where pronounced deviations from the predicted values are found \cite{Zeman1999}.

\section{CONCLUSION}
From ultrasonic measurements in hydrogenated and deuterated (TMTTF)$_2$PF$_6$ salts, we observed magneto-elastic coupling effects extending over a wide temperature range beyond the CO phase transition. In the hydrogenated salt, the relaxation dependent behavior of the compressibility modulus $C_{33}$ is found to disclose highly inhomogeneous character of the non ordered SP phase, most likely due to the proximity of the SP and CO phase transitions in this material and to the amplitude of the spin-lattice coupling that is apparently much weaker compared to the deuterated one. For both compounds, there is an upturn in the modulus stiffening below 40 K, which is correlated to the loss of the spin degrees of freedom observed in numerous magnetic experiments. An anomaly is clearly observed at the SP transition for the deuterated salt only. Although the SP phase appears only weakly inhomogeneous, the elastic anomaly at $T_{\rm SP}$ is consistent with a biquadratic coupling between the elastic deformation and the SP order parameter, a coupling that decreases in a magnetic field. Precursor softening effects at $T_{\rm SP}^*$ are rather explained by a linear-quadratic coupling. Despite the inhomogeneous phase, the magnetic field dependence of the SP transition temperature agrees with the mean-field predictions.

\acknowledgments{The authors acknowledge the technical support of Mario Castonguay. This work was supported by grants from the Fonds Qu\'eb\'ecois de la Recherche sur la Nature et les Technologies (FQRNT) and from the Natural Science and Engineering Research Council of Canada (NSERC).}


\begin{thebibliography}{23}
\bibitem{Bourbon2008} C. Bourbonnais and D. J\'erome,  in {\it  Physics of organic Superconductors and Conductors}, edited by A. G. Lebed, Vol. 110, Springer Series in Materials Science (Springer, Heidelberg, 2008), P. 357.

\bibitem{Brown2008} S. E. Brown, P. M. Chaikin, and M. J. Naugthon in {\it  Physics of organic Superconductors and Conductors}, edited by A. G. Lebed, Vol. 110, Springer Series in Materials Science (Springer, Heidelberg, 2008), P. 49.

\bibitem{Monceau2008} P. Monceau, J.-C. Lasjaunias, K. Biljakovic, and F. Nad in {\it  Physics of organic Superconductors and Conductors}, edited by A. G. Lebed, Vol. 110, Springer Series in Materials Science (Springer, Heidelberg, 2008), P. 277.

\bibitem{Creuzet1985} F. Creuzet, Mol. Cryst. Liq. Cryst. \textbf{119}, 289 (1985).

\bibitem{Caron88} L.G. Caron, F. Creuzet, P. Butaud, C. Bourbonnais, D. J\'erome and K. Bechgaard, Synth. Met. {\bf 27}, 123 (1988).

\bibitem{Chow98} D. S. Chow,  P. Wzietek,  D. Fogliatti, B. Alavi,  D. J. Tantillo,  C. A. Merlic,  and S. E. Brown, Phys. Rev. Lett. {\bf 81}, 3984 (1998).

\bibitem{Creuzet1987} F. Creuzet, C. Bourbonnais, L.G. Caron, D. Jerome and K. Bechgaard, Synth. Met. \textbf{19}, 289 (1987).

\bibitem{Dumm00} M. Dumm, A. Loidl, B. W. Fravel, K. P. Starkey, L. K. Montgomerey, and M. Dressel, Phys. Rev. B {\bf 61}, 511 (2000).

\bibitem{Foury08} P. Foury-Leylekian, S. Petit, C. Coulon, B. Hennion, A. Moradpour, and J.-P. Pouget, Physica B \textbf{404}, 537 (2009).

\bibitem{Wzietek1993} P. Wzietek,F. Creuzet, C. Bourbonnais, D. J\'erome, K. Bechgaard and P. Batail, J. Phys. I \textbf{3}, 171 (1993).

\bibitem{Zamborszky2002B} F. Zamborszky, W. Yu, W. Raas, S. E. Brown, B. Alavi, C. A. Merlic, and A. Baur, Phys. Rev. B {\bf 66}, 081103(R) (2002).

\bibitem{Brown1998} S. E. Brown, W. G. Clark, F. Zamborsky, B. J. Klemme, G. Kriza, B. Alavi, C. A. Merlic, P. Kuhns and W. Moulton, Phys. Rev. Lett. \textbf{80}, 5429 (1998);  S. E. Brown, W. G. Clark, B. Alavi, M. J. Naugthon., D. J. Tantillo  and C. A. Merlic, Phys. Rev. B \textbf{60}, 6270 (1999).

\bibitem{Pouget1982} J. P. Pouget, R. Moret, R. Comes, K. Bechgaard;, J. M. Fabre and L. Giral, Mol. Cryst. Liq. Cryst. \textbf{79}, 485 (1982).

\bibitem{Foury04} P. Foury-Leylekian, D. Le Bolloc'h, B. Hennion, S. Ravy, A. Moradpour, and J.-P. Pouget, Phys. Rev. B {\bf 70}, 180405 (2004).

\bibitem{Pouget2006} J.-P. Pouget, P. Foury-Leylekian, D. Le Booloc'h, B. Hennion, S. Ravy, C. Coulon, V. Cardoso and A. Moradpour, J. Low Temp. Phys \textbf{142}, 147 (2006).

\bibitem{Bourbonnais96} C. Bourbonnais and B. Dumoulin, J. Physique I (France) {\bf 6}, 1727 (1996)

\bibitem{deSouza2008} M. de Souza, P. Foury-Leylekian, A. Moradpour, J.-P. Pouget and M. Lang., Phys. Rev. Lett. \textbf{101}, 216403 (2008).

\bibitem{deSouza2009} M. de Souza, A. Br$\ddot{u}$hl, J. M$\ddot{u}$ller, P. Foury-Leylekian, A. Moradpour, J.-P. Pouget and M. Lang, Physica B \textbf{404}, 494 (2009).

\bibitem{Langlois2010} A. Langlois, M. Poirier, C. Bourbonnais, P. Foury-Leylekian, A. Moradpour and J.-P. Pouget, Phys. Rev.B \textbf{81}, 125101 (2010).

\bibitem{Cross1979} M.C. Cross, Phys. Rev. B \textbf{20}, 4606 (1979).

\bibitem{Nad2000} F. Nad, P. Monceau, C. Carcel and J.M. Fabre, Phys. Rev. B \textbf{62}, 1753 (2000).

\bibitem{Coulon82} C. Coulon,  P. Delhaes,  S. Flandrois,  R. Lagnier,  E. Bonjour, and J.M. Fabre, J. Phys. (Paris) {\bf 43} 1059 (1982).

\bibitem{Brown1989} S.E. Brown, H.H.S. Javadi and R. Laversanne, MRS Proceedings, \textbf{173}, 245 (1989).

\bibitem{Trudeau1992} Y. Trudeau, M. Poirier and A. Caill\'{e}, Phys. Rev. B \textbf{46}, 169 (1992).

\bibitem{Poirier1995} M. Poirier, M. Castonguay, A. Revcolevschi and G. Dhalenne, Phys. Rev. B \textbf{52}, 16058 (1996).

\bibitem{Dumoulin1997} B. Dumoulin, P. Fronzse, M. Poirier, A. Revcolevschi and G. Dhalenne, Synthetic Metals \textbf{86}, 2243 (1997).
\bibitem{Laversanne84} R. Laversanne,  C. Coulon, B. Gallois, J. P.
Pouget and R. Moret, J. Phys. (Paris)  Lett. {\bf 45}, L393 (1984).

\bibitem{Pouget1989} J. Pouget, in \textit{Low-Dimensional Elctronic Properties of Molybdenum Bronzes and Oxides}, edited by C. Schlenker (Kluwer Academic, Netherlands, 1989), pp. 87-157.

\bibitem{Azzouz1996} M. Azzouz and C. Bourbonnais, Phys. Rev. B \textbf{53}, 5090 (1996).

\bibitem{Bloch1980} D. Bloch, J. Voiron, J.C. Bonner, J.W. Bray, I.S. Jacobs, and L.V. Interrante, Phys. Rev. Lett. \textbf{44},294 (1980).

\bibitem{Zeman1999} J. Zeman, G. Martinez, P.H.M. van Loosdrecht, G. Dhalenne and A. Revcolevschi, Phys. Rev. Lett. \textbf{83}, 2648 (1999).



\end{thebibliography}
\end{document}